\documentclass{aip-cp}

\usepackage[numbers]{natbib}
\usepackage{rotating}
\usepackage{graphicx}

\begin{document}

% Title portion
\title{Scaling Properties of the Lipkin Model at the Critical Point}

\author[aff1,aff3]{J.M. Arias\noteref{note1}}
\eaddress{ariasc@us.es}

\author[aff2,aff3]{J.E. Garc\'{\i}a-Ramos\noteref{note1}}
\eaddress{enrique.ramos@dfaie.uhu.es}

\affil[aff1]{Departamento de F\'{\i}sica At\'omica, Molecular y Nuclear, Facultad de F\'{\i}sica, Universidad de Sevilla, Apartado~1065,
  41080 Sevilla, Spain}

\affil[aff2]{Departamento de  Ciencias Integradas y  Centro de Estudios  Avanzados en F\'isica,   Matem\'atica y Computaci\'on, Universidad de Huelva, 21071 Huelva, Spain}

\affil[aff3]{Instituto Carlos I de F\'{\i}sica Te\'orica y Computacional, Universidad de Granada, Fuentenueva s/n, 18071 Granada, Spain}

\authornote[note1]{Dedicated to Professor Francesco Iachello as recognition of all his career with our profound gratitude for his continuous advice and support.}
%\authornote[note2]{This is an example of second authornote.}
%\corresp[cor1]{Corresponding author: enrique.ramos@dfaie.uhu.es}

\maketitle

\begin{abstract}
The influence of Franco Iachello in Physics during the last 50 years and, in particular, in the use of algebraic methods applied to very different physical problems has been broad, extense and have permeated most branches of Physics, from Nuclear and Molecular to Particle and Condensed Matter physics. Apart of many other contributions, at the beginning of the 2000's he introduced the concept of critical point symmetry and triggered the study of the many faces of quantum phase transitions in nuclei and other mesoscopic systems. In this contribution, we present the analysis of the scaling properties of the Lipkin model spectrum at the phase transition point and we focus on the differences between first and second order quantum phase transitions. Moreover, we explain how the obtained results can be also of application for the interacting boson model.
\end{abstract}

\section{INTRODUCTION}
Thermodynamic phase transitions are phenomena that can be found everywhere in Nature and they involve sudden changes in the structure of the matter. For example, when water passes from its liquid phase to ice at $0^\circ$C, there is an abrupt change in its structure, going from a disordered phase to an ordered one. Moreover, the density suffers a sudden change decreasing roughly a $10\%$ and the specific heat shows a discontinuity, too.  Another well known example appears in magnetic materials, such as iron, when it passes from its diamagnetic to its ferromagnetic phase at the Curie temperature, $T_c=1043$ K. In this case, the derivative of the magnetization of the substance presents a discontinuity. In both cases,  a control parameter exists, namely, the temperature, which leads the system into one or another phase. Besides, an order parameter (the density in the first case and the magnetization in the second), which defines the phase, either ordered of disordered, can be also defined. The first case corresponds to a first order phase transition while the second, to a second order one, also known as continuous phase transition. Thermodynamic phase transitions are characterized by a set properties that are scale invariant. This special behavior was nicely explained through {\it the scaling hypothesis}  in the theory of critical exponent of Widom and Kadanov \cite{Kada67}. The properties of the system at the critical point become independent of the fine characteristic of the interaction between the elementary components of the system and, therefore, are universal.  Indeed, one can assign every system to a class of universality that mostly depends on its dimensionality. Two systems belonging to the same universality class will possess the same critical exponents and therefore the same thermodynamic properties at the critical point.

In recent years a growing interest in the study of a new kind of phase transition, called Quantum Phase Transitions (QPT), has been developed. The concept of QPT was treated in one of the first occasions in the seminal work by Gilmore and Feng \cite{Gilm77} where they studied the Lipkin model in its pseudospin form, although they called to this phenomenon {\it ground state energy phase transition}. In this work, the authors defined in a precise way the potential energy for the ground state, which depends on a set of order and control parameters and they analyzed the behavior of the stationary points of the potential as a function of the control parameters through the use of the Catastrophe Theory \cite{Gilm81}. This analysis, that corresponds to the thermodynamic limit, i.e.,  the size of the system goes to infinity, allowed to find the appearance of a second order QPT in the Lipkin model.  
QPTs are different from the thermodynamic phase transitions in several aspect, e.g., they happen at zero temperature, the system can have a finite size, and the control parameter is not simply the temperature or the pressure but one (or more) parameter that corresponds to a relative strength ($x$) between two adding parts in the system Hamiltonian, having each of these two parts their own symmetry. Therefore, the Hamiltonian can be written as,
\begin{equation}
H=x H_1+(1-x)H_2 ,
\end{equation}
where $H_1$ has the symmetry $1$ while $H_2$ has the symmetry $2$. The concept of QPT was promptly applied to the Interacting Boson Model (IBM) \cite{IBM} in a set of almost simultaneous works \cite{DSI80,Gino80a,Gino80b,FGD}, where the authors found the existence of a first order QPT line and a second order QPT point in the model. To such an end, a mean-field approach was used, introducing the so called intrinsic/coherent state which was defined in terms of a condensate of bosons that depends on two variational parameters that characterize the shape of the nucleus. 

The use of a coherent state supposes the analysis of the QPT in the large size limit, i.e., in the thermodynamic limit. However, in many situations it is needed to study the problem for a finite system, i.e., for a finite number of particles, $N$. In particular, the nucleus is a finite system and, therefore, a finite number of particle, nucleons, should be considered. Even in such situations, some precursors of the phase transitions can be observed \cite{Iach04}, namely, rapid changes in the derivative of the ground state energy and in the value of the order parameter. Therefore, it is of great interest to know how the properties of the system evolve at the critical point as one moves to the thermodynamic limit. This evolution is described through the finite size scaling exponents \cite{Fish72}, which provide the information on how different observables of the system evolve as N tends to infinity in second order (or continuous) phase transitions. 

In this contribution, we review certain features of QPTs in the Lipkin model, focusing on the scaling properties of the spectrum in terms of the system size and on the differences between first and second order QPTs. 

\section{THE MODEL}
The Lipkin model was proposed by Lipkin, Meshkov, and Glick \cite{Lipk65} and in its boson realization it can be written in terms of scalar bosons that can occupy two non-degenerated energy levels labeled by $s$ and $t$. Therefore, the dynamical algebra of the system is $u(2)$, which means that the Hamiltonian can be expressed in terms of four possible bilinear products of one creation, $s^\dag$ and  $t^\dag$, and one annihilation, $s$ and $t$, boson operator. In this work we will focus on a restricted version of the Hamiltonian called Consistent-Q Hamiltonian (CQF), which resembles the used one in the IBM \cite{CQIBM}.

The Hamiltonian of the model can be written as, 
\begin{equation}
  \label{eq:hamiltonian}
  H=x~  n_t-\frac{1-x}{N} ~ Q^{y} Q^{y},
\end{equation}
where the operators $n_t$ and $Q^{y}$ are defined as
\begin{equation}
  \label{eq:Qdef}
 n_t=  t^\dag t, \qquad Q^{y}= s^\dag s +t^\dag t +y \left (d^\dag d \right).
\end{equation}
The two parameters $x$ and $y$ correspond to the two control parameters of the system (the introduction of the $y$-term allows for a first order phase transition in the Lipkin model). 

To obtain the behavior of the system in the thermodynamic limit we use a boson condensate to generate the mean-field energy
\begin{equation}
|g\rangle=\frac{1}{\sqrt{N!}}\left(\frac{1}{\sqrt{1+\beta^2}}~(s^\dag+\beta t^\dag)\right)^{N}|0\rangle, 
\end{equation} 
where $\beta$ is a variational parameter, $|0\rangle$ is the boson vacuum and $N$ the number of bosons. The energy surface in the large $N$ limit is written down as,
\begin{eqnarray}
\label{ES}
E(N,\beta)&=&\langle N,\beta  | H | N,\beta \nonumber \rangle\\
&=& N {\beta^2 \over (1+\beta^2)^2} \Big\{ 5x-4+4 \beta  y (x-1)+ \beta^2\big[x+y^2(x-1) \big] \Big\} .
\end{eqnarray}

We would like to emphasize that the energy (\ref{ES}) coincides with that of the CQF IBM Hamiltonian under the assumption of axial symmetry, which implies that both models own the same phase diagram, but not necessarily the same properties related to the spectra. The simplicity of the Lipkin model compared with the IBM allows the numerical analysis (through a direct diagonalization) of QPTs with much larger number of particles.

The study of the energy surface (\ref{ES}) allows to obtain the position of the QPT points as a function of the control parameters, $x\in [0,1]$ and $y \geq 0$, which corresponds to $x_c=(4+y^2)/(5+y^2)$. In the case of $y=0$, the observed QPT is of second order while for $y\neq 0$ the QPT is of first order. The system can only present two possible phases, either  deformed with $\beta_0\neq 0$ for $x<x_c$, or spherical with $\beta_0=0$ for $x>x_c$.

\section{THE SCALING PROPERTIES}
\label{sec-scaling}
The presence of a QPT supposes the appearance of discontinuities, or sudden changes in the case of finite systems, in many properties of the quantum systems. In particular, the excitation energy of the first excited state, also known as gap, presents a rapid change around the critical point. The focus of this section will be on the study of the behavior of the gap around the critical point, emphasizing the differences between first and second order QPTs.
\begin{figure}
\includegraphics[width=0.5\linewidth]{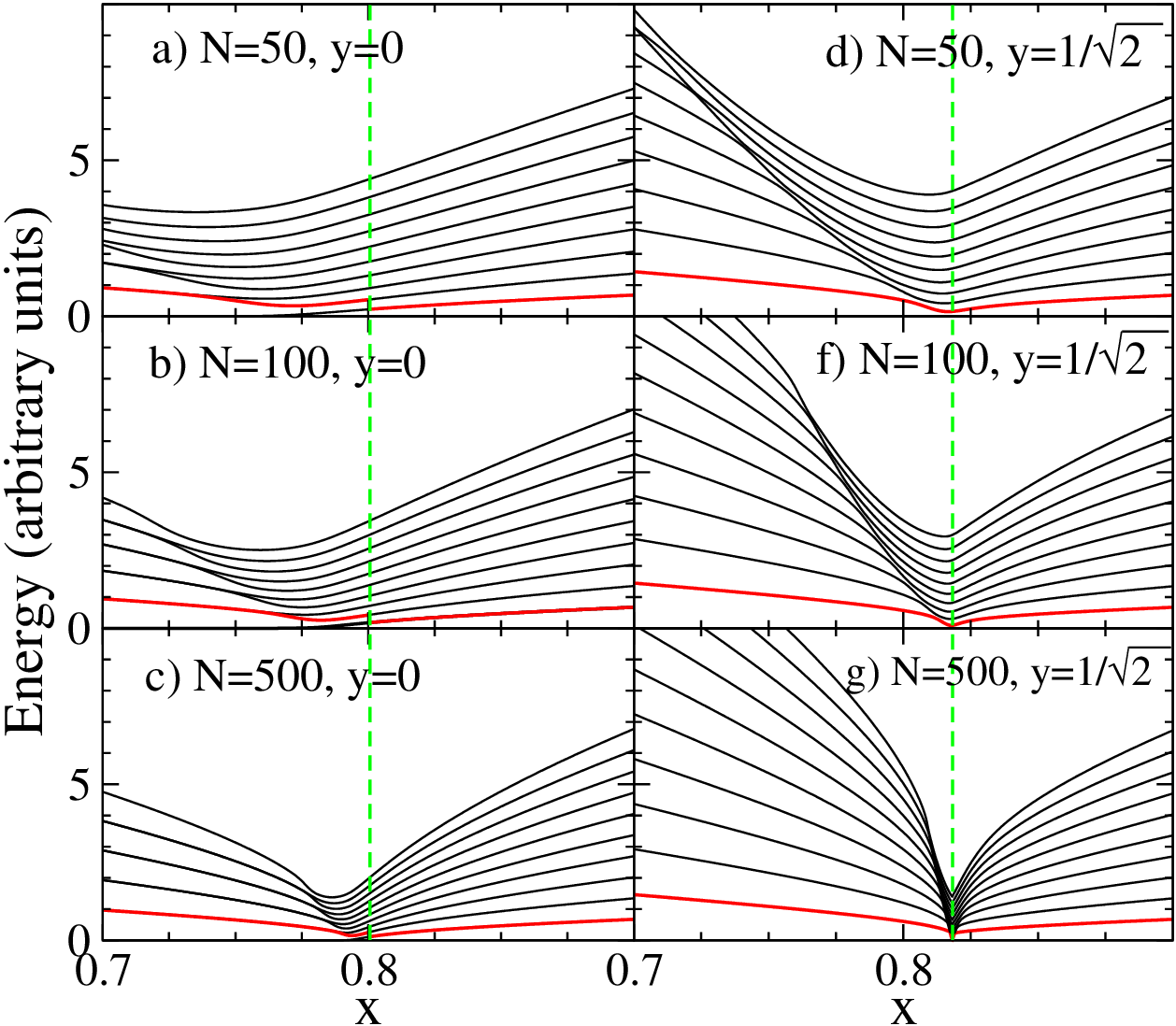}
%\vspace{1cm}
%\includegraphics[width=1.0\linewidth]{fig_schdic.eps}
\caption{Excitation energies, for the Lipkin model described by Eq.~(\ref{eq:hamiltonian}), of the first ten excited states as a function of the control parameter, $x$, for different values of the number of particles, $N=50$, $100$ and $500$ and two values of $y$, $y=0$ and $y=1/\sqrt{2}$. Red curve corresponds to the excitation energy of the first excited state (see text for details), and the green dashed line to the position of the critical point in the thermodynamic limit, namely, $x_c=4/5$ for $y=0$ and $x_c=9/11$ for $y=1/\sqrt{2}$.}
\label{fig-spectra-gap}
\end{figure}

Scaling properties, are usually written as
\begin{equation}
\Xi(N,\lambda_c)= A\, N^\theta ,
\label{eq-power}
\end{equation}
stating that an observable $\Xi(N,\lambda_c)$, function of the system size ($N$) behaves at the critical point, $\lambda_c$, as a universal power $\theta$  (critical exponent) of $N$. Stricktly speaking Eq.~(\ref{eq-power}) can only be defined in second order QPTs. In other words, the term {\it critical point} can only be properly used in the case of second or higher order QPTs, though in the Nuclear Physics realm it is widely used for first order QPTs as well. 

To appreciate how the spectrum of the Lipkin model behaves differently in first and second order QPT, we present in Fig.~\ref{fig-spectra-gap} the first ten excitation energies for a number of bosons $N=50, 100$ and $500$ and $y=0$ for the second order QPT (panels a), b), and c)) and $y=1/\sqrt{2}$  for the first order one (panels d), f), and g)), as a function of the control parameter, $x$, around the critical point. These are $x_c=4/5$ for the second order QPT, and $x_c=9/11$ for the first order QPT. It is easily observed how the spectra are compressed around the critical point, being more stretched in the case of the first order QPT compared with the second order one. Moreover, it is also noticeable how the larger the number of bosons, the smaller the gap is. The way the gap (red curve in Fig.~\ref{fig-spectra-gap}) behaves is different in second and first order QPTs. Note that in the case with $y=0$ the Hamiltonian preserves the parity, with states with either an even or an odd number of $t$ bosons. In this case and in the deformed phase ($x<x_c$), the states are degenerated in couples, while the degeneration is broken in the spherical phase ($x>x_c$). Therefore, the gap is defined in the second order QPT as the excitation energy of the first excited state having the same parity than the ground state, and, as a matter of fact, it implies that the corresponding state should change from the second, in the deformed sector, to the first excited one in the spherical phase. In the case of first order QPT, the gap always corresponds to the excitation energy of the first excited state. An extra element that can be readily observed in the spectrum is the line of large state density that appears to the left of the critical point, known as Excited State Quantum Phase Transition (ESQPT). This concept was introduced by Caprio, Cejnar and Iachello in \cite{Capr08}.

%In deep
The rigorous form to obtain the finite size scaling exponents in the case of second order QPTs is to analyze the spectrum of the Hamiltonian under study in the large $N$ limit. This task is carried out through a $1/N$ expansion, using the Holstein-Primakoff boson representation of the Hamiltonian followed by, either  a Bogoliubov transformation, that is only valid to diagonalize the Hamiltonian at $1/N^0$ order \cite{Aria07}, or a continuous unitary transformations to diagonalize the Hamiltonian at every order in $1/N$ \cite{Dusu04}. The key result of this analysis is that every observable at the critical point can be described through an expression with a regular and a singular part:
\begin{equation}
\Phi_N(\lambda)=\Phi_N^{reg}(\lambda)+\Phi_N^{sing}(\lambda).
\end{equation}  
Taking into account that the behavior of the observables cannot diverge at the critical point and following the ideas developed in Ref.~\cite{Fish72}, one can assume that the diverging part of the function should behave as a scaling function.
To obtain  the value of the finite size scaling exponent one has simply to analyze $\Phi_N^{sing}$ \cite{Dusu04,Dusu05,Dusu05b}. Another clever way of getting the finite size scaling exponent appears in Ref.~\cite{Rowe04} where the authors showed that at the critical point the whole spectrum scales as $N^{-1/3}$. Indeed, as it was shown in \cite{Dusu05b}, the scaling exponent of the gap, $-1/3$, is not only valid for the Lipkin model, but universal for any other two-level model, as for example the IBM or the vibron model. A similar study can be carried out for other observables obtaining the corresponding scaling exponents (see \cite{Dusu05b} for details).
\begin{figure}
\includegraphics[width=0.5\linewidth]{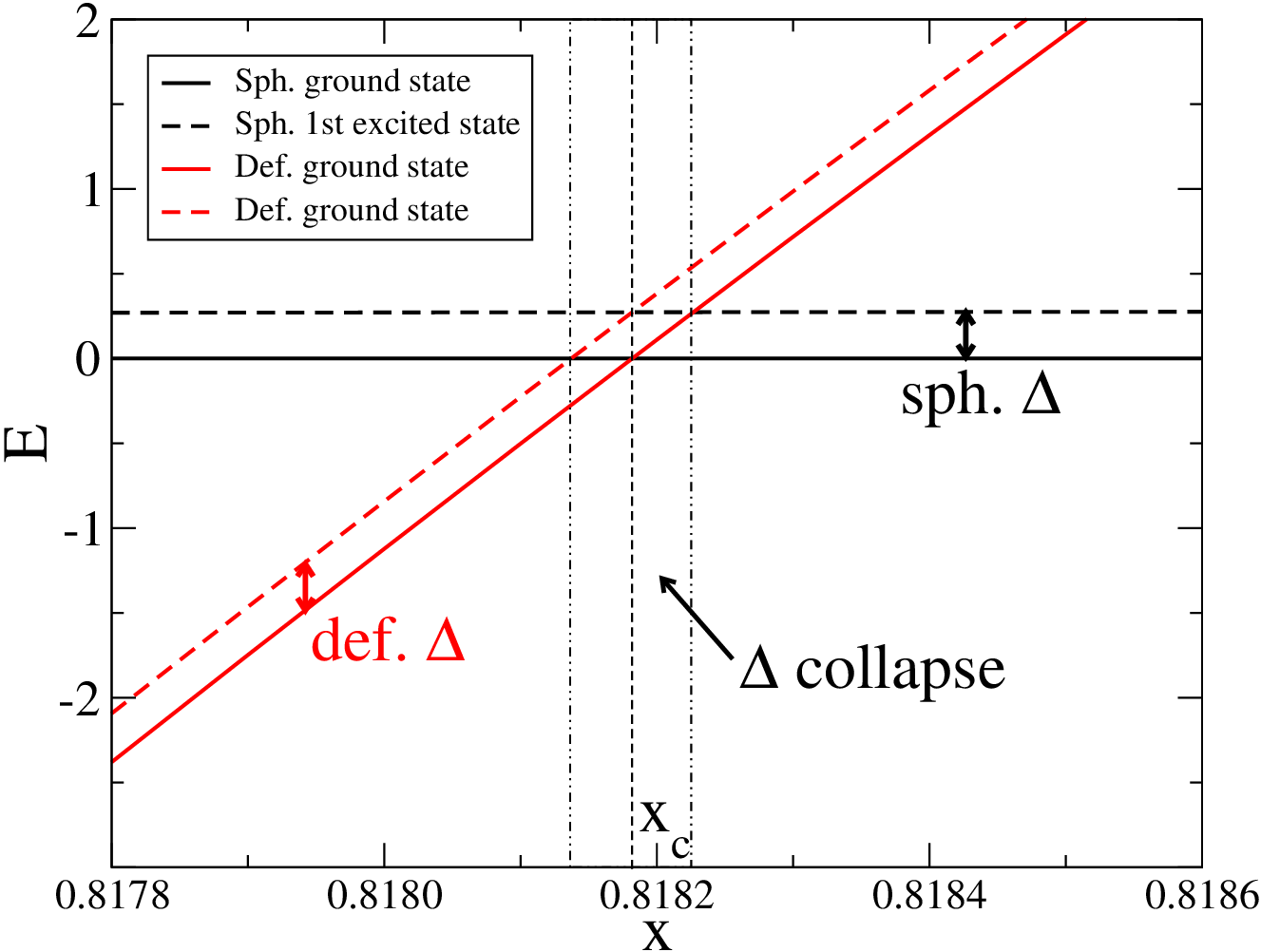}
\caption{Mean field ground state energies and gaps of the spherical and the deformed phases for the Lipkin model with N=10000  around the critical point, $x_c=9/11$.}
\label{fig-gap1}
\end{figure}

In the case of first order QPTs, the analysis is rather different. In particular, in Ref.~\cite{Rowe04} it was clearly established how the scaling argument used for a second order QPT is no longer valid in the case of first order and, therefore, it cannot be shown whether a finite size scaling law exists or not. It is worth to stress that the first order critical point arises from the crossing of two sets of states with very different structures, e.g., one spherical (symmetric) while other deformed (broken). Note that in this case there exists a narrow coexistence region around the first order critical point. In Fig.~\ref{fig-gap1} we plot, for the Lipkin model, the absolute mean field energy of the spherical and the deformed minima for $y=1/\sqrt{2}$ as a function of $x$, in the coexistence region, together with the energy of the first excited state of each configuration. At the left hand side of the figure, the ground state of the system is deformed and the first excited state also corresponds to a deformed state. As one moves into the $x_c$ point, one first crosses the vertical left dotted-dashed line and, although, the ground state is still deformed, the first excited state corresponds to the spherical ground state. At $x_c$ (dashed line), the deformed and the spherical ground states become degenerated and the gap vanishes. From this point up to the right vertical dotted-dashed line, the ground state is spherical while the first excited state corresponds to the deformed ground  state. From here onwards, the ground and the first excited state are both spherical. Note that the region between both vertical dotted-dashed line is extremely tiny, $\Delta x\approx 5\cdot 10^{-5}$. Therefore, the gap has a rather smooth behavior in the whole range of the control parameter except in a very narrow region around the first order QPT point, where it vanishes. In the $N$ infinite limit, this narrow region collapsess into a single point.

\begin{figure}
\includegraphics[width=0.5\linewidth]{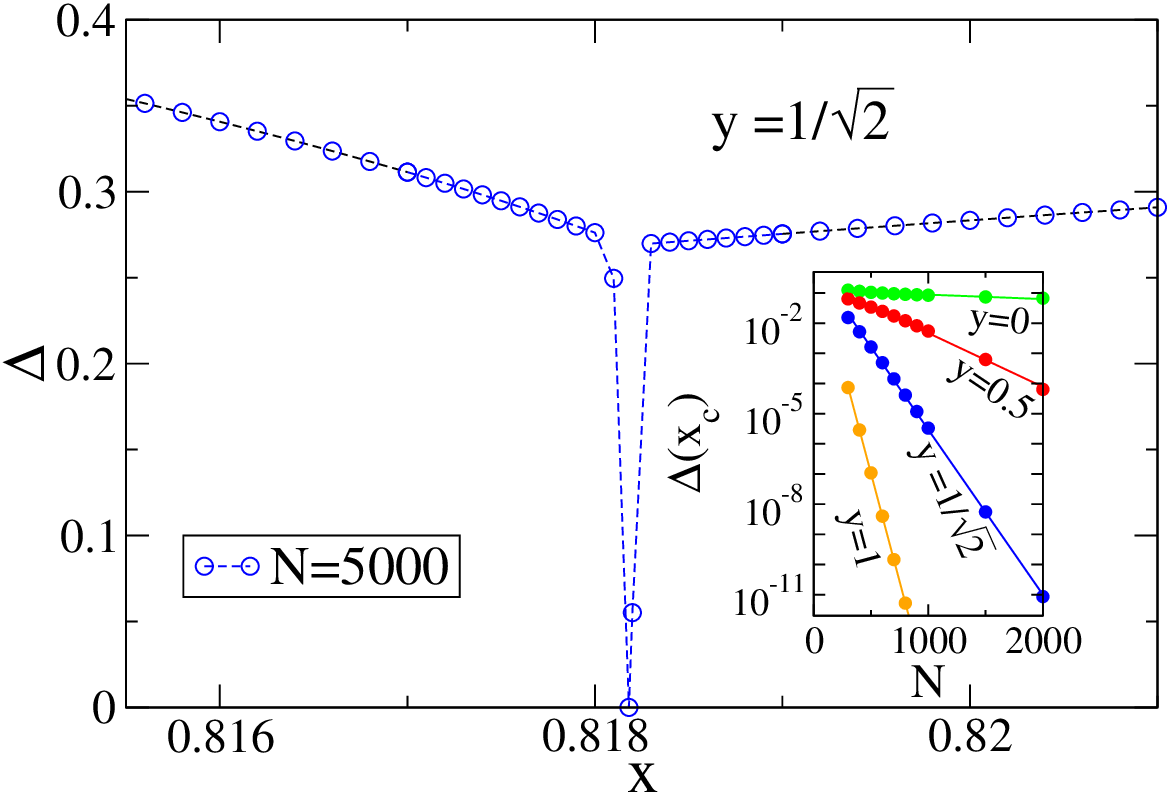}
\caption{Numerical value of the gap in the Lipkin model for $y=1/\sqrt{2}$  and $N=5000$, around the critical point,
  $x_c=9/11$. In the inset it is plotted the exponential decrease of the gap at the corresponding 
  critical point for several values of $y$. Figure adapted from Ref.~\cite{Vida06}.}
\label{fig-gap2}
\end{figure}

In Fig.~\ref{fig-gap2} it is plotted the value of the gap in the Lipkin model around the first order critical point for $N=5000$ and $y=1/\sqrt{2}$. The gap at the left and right of the first order critical point, $x=x_c^{\pm}$, are equal and finite but vanishes at $x=x_c$, as it has been shown in Fig~\ref{fig-gap1}, i.e., this is a direct consequence of the crossing of two different ``ground'' states at the first order QPT point. As a consequence, the scaling power law of the gap is no longer valid and the dependence of the gap with the size of the system at the critical point follows an exponential law, $\Delta\sim e^{-A N}$ \cite{Vida06}. This is clearly shown in the inset of Fig.~\ref{fig-gap2} where it is fitted the value of the gap for different values of $y$ with an exponential function. 
This special behavior is particular for the Lipkin model, where the $1/N^i$ corrections are equal for every order, $i$, for the spherical and the deformed phases. The case of the IBM is different and the later is not fulfilled. As a matter of fact, Fig.~\ref{fig-gap1} is  still valid for IBM: the mean field ground states, spherical and deformed, become degenerated at $x_c=9/11$ and the two excited states also cross at the same point. However, due to the angular momentum content of the IBM, $\Delta$ does not collapse anymore. On one hand, the gap corresponds to the band-head of the $\beta$ band in the deformed phase (not to the first $2^+$ state), and to the one phonon state ($2_1^+$), in the spherical phase, while the spherical and the deformed ground states both correspond to  $0^+$ states, as also happened in the Lipkin case. Therefore, at the critical point, once the ground states cross, the gap will correspond to the excitation energy of the first state with angular momentum  $L=2$. At  $x_c=9/11$, the gap can be easily calculated using Eq.~(18) of \cite{Aria07}. $\Delta$ is therefore finite in the $N$ infinite limit with a value $\Delta=3/11$  which is the same that for the Lipkin model at $x=x_c^{\pm}$ (see Fig.~\ref{fig-gap2}). This is in clear contradiction with what it was claimed in \cite{Will10}, where the authors established that there was an universal scaling behavior in first order QPTs, as it happens for second order ones, which is not true.   

\section{CONCLUSIONS}

In summary, the universal scaling power law (\ref{eq-power}) is only valid for second order QPTs and in this case the whole spectrum scales at the critical point with the law $N^{-1/3}$, although other observables can have different finite size scaling exponents. This behavior is the same for the Lipkin, the IBM and any other two-level boson model. In the case of a first order QPT, there is no universality and it does not exist any finite size scaling exponent. In the case of the Lipkin model, the dependence of the gap with the system size at the first order critical point follows an exponential law while in the case of the IBM, the gap goes asymptotically to a finite value. Finally, note that the way the different energies decrease as $N$ increases at the first order phase critical point is not universal as can be readily observed from Table I of Ref.~\cite{Will10}, i.e., the fitted scaling parameters change as a function of the angular momentum, except for the second order QPTs, where the universality is recovered.

\section{ACKNOWLEDGMENTS}
This work has been supported by the Spanish Ministerio de Econom\'{\i}a y Competitividad and the European regional development fund (FEDER) under Projects No. FIS2017-88410-88410-P and by Consejer\'{\i}a de Econom\'{\i}a, Innovaci\'on, Ciencia y Empleo de la Junta de Andaluc\'{\i}a (Spain) under Group FQM-160 and FQM-370.
We also want to thank Franco Iachello for his friendship and for inspiring us during so many years in our studies in Nuclear and Molecular Physics, in general, and in the field of Quantum Phase Transitions, in particular. 
% References

\nocite{*}
\bibliographystyle{aipnum-cp}%
%\bibliography{sample}%

\end{document}